\documentclass{elsart}          %% LaTeX 2e (preferred)
\usepackage{amstext}
\usepackage{amsxtra}
\usepackage{amssymb}
\usepackage{graphicx}

\journal{Optics Communications}

\begin{document}
\begin{frontmatter}
\title{Interference-filter-stabilized external-cavity diode lasers}

\author{X. Baillard,}
\author{A. Gauguet,}
\author{S. Bize,}
\author{P. Lemonde,}
\author{Ph. Laurent,}
\author{A. Clairon,}
\author{P. Rosenbusch}
\author{}
%% for REVTeX4, each author name can be set in a separate \author{} field

\address{Syst\`emes de R\'ef\'erence Temps-Espace$^{*}$, Observatoire de Paris,
\\  61 avenue de l'Observatoire, 75014 Paris, FRANCE}

\ead{Peter.Rosenbusch@obspm.fr} \ead[url]{opdaf1.obspm.fr}

\begin{abstract}We have developed external-cavity diode lasers,
where the wavelength selection is assured by a low loss interference
filter instead of the common diffraction grating. The filter allows
a linear cavity design reducing the sensitivity of the wavelength
and the external cavity feedback against misalignment. By separating
the feedback and wavelength selection functions, both can be
optimized independently leading to an increased tunability of the
laser. The design is employed for the generation of laser light at
698, 780 and 852~nm. Its characteristics make it a well suited
candidate for space-born lasers.
\end{abstract}

\end{frontmatter}

\section{Introduction}

Semiconductor lasers have become an inexpensive easy-to-handle
source of coherent light. Applications in many fields such as atomic
physics, metrology and telecommunication require single mode
operation with narrow linewidth and good tunability \cite{Wieman}.
This is frequently achieved by incorporating the laser diode into an
external cavity where optical feedback and wavelength discrimination
are provided by a diffraction grating \cite{Littrow}. However, such
a design is sensitive to the ambient pressure
\cite{indexOfRefraction} and to optical misalignment
\cite{OpticsLetters} induced by mechanical or thermal deformation.
In addition, for the common Littrow configuration, the direction or
position of the output beam depends on the wavelength
\cite{Haensch,Australia}. An alternative design employs a
Fabry-P\'erot etalon as wavelength discriminator, operating in
transmission \cite{SYRTErelock}. In that case the two tasks of
wavelength selection and feedback reflection are carried-out by two
different optical elements. The etalon can either be formed from a
thin air gap between two glass plates or a single, thin ($\sim30
\mu$m) solid plate. Both solutions are costly, can be fragile and
exhibit multiple resonances. Furthermore, we observe that the
absorption of atmospheric water can shift the transmitted wavelength
to the point of uselessness after two years.

Better robustness and unique transmission is provided by narrow-band
dielectric interference filters. Their use in an extended cavity
laser has been demonstrated for telecom wavelength in
\cite{OpticsLetters}. The emission from the anti-reflection coated
back facet of a $\lambda=1300$~nm laser diode is fed-back by a
"cat's eye" (lens + mirror). An interference filter of 2~nm passband
width and 70~\% peak transmission acts as intra-cavity wavelength
discriminator.

%The diode's output facet serving as outcoupler does not seems to
%have any particular coating thereby fixing the reflection
%coefficient to 25\% \cite{indexGaAS}.

Here, we present filters having 90~\% transmission and $\sim0.3$~nm
FWHM at near infra-red and visible wavelengths. The external cavity
is added to the output beam, while the diode's back facet is coated
for high reflectivity. The cavity outcoupler is a partially
reflecting mirror. Changing its reflectivity is an easy way of
optimizing the feed-back. The diode's output facet has no particular
high-quality anti-reflection coating giving rise to a second cavity
formed by the laser chip itself, but making it inexpensive. Chaotic
coupling to adjacent chip modes is efficiently suppressed by the
narrow bandwidth of our filter, up to 8 times the threshold current.

 We show that the sensitivity of these lasers to
environmental perturbations is drastically reduced as compared to
the Littrow configuration. We demonstrate tunability over a broader
wavelength range thanks to the possibility of optimizing the amount
of feedback independently from the wavelength selection mechanism.
We study the lasers' spectral noise.

\section{The external cavity}

\begin{figure}[t]
\centerline{\includegraphics{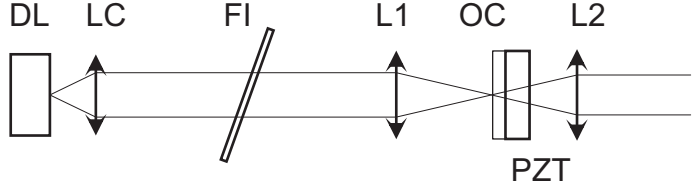}}
\caption{Schematic of the external cavity laser using an
interference filter (FI) for wavelength selection: (DL) laser diode,
(LC) collimating lens, (OC) partially reflective out-coupler, (PZT)
piezoelectric transducer actuating OC, (L1) lens forming a "cat's
eye" with OC, (L2) lens providing a collimated output beam.
\label{schematic}}
\end{figure}

A schematic of the external cavity is given in Fig.
\ref{schematic}. The light emitted from the diode (DL) is
collimated by an objective lens (LC) with short focal length (3 to
4.5 mm) and high numerical aperture ($\sim0.6$). The lens is
chosen to compensate for abberations arising from the diode's
packaging window. A partially reflecting mirror, here named
out-coupler (OC) provides the feedback into the diode. The OC is
displaced by a  piezo-electric transducer (PZT) in order to vary
the cavity length. A narrow-band high-transmission interference
filter (FI) is introduced into the cavity. The filter provides the
frequency selectivity usually obtained by replacing the
out-coupler with a diffraction grating. With this set-up, we are
able to achieve single-mode, tunable operation. In addition much
better stability against optical misalignment is achieved by
focussing the collimated beam in a "cat's eye" onto the
out-coupler. We typically employ a lens (L1) of 18~mm focal
length. A second, similar lens (L2) provides a collimated output
beam. Contrary to the Littrow laser design, reflection and
wavelength discrimination are provided by two different elements
so that the amount of feedback can easily be optimized.

In Fig. \ref{schematic} the diode's back facet is coated for high
reflection. Alternatively, this facet can have low reflection to
provide the output beam. In that case one chooses OC with maximum
reflectivity.

\section{Tunability and wavelength sensitivity}

For a Littrow laser wavelength discrimination is given by the Bragg
condition $\lambda=2d\sin\theta$ where $d$ is the grating's line
spacing and $\theta$ the angle of incidence. For wavelengths in the
near infra-red, typical values are $d^{-1}=1200$~ lines/mm and
$\theta=30^{\circ}$. This leads to
$d\lambda/d\theta\approx1.4\;$nm/mrad. Tuning the laser redirects
the output beam by $2(d\lambda/d\theta)^{-1}\approx1.4\;$mrad/nm or,
if a mirror is attached to the grating mount\cite{Australia}, leads
to a transverse displacement of $dx/d\lambda=18\;\mu$m/nm (assuming
a distance of 15~mm between the grating and the mirror).

Wavelength discrimination of our filter is based on multiple
reflection within its dielectric coatings and behaves as a thin
Fabry-P\'erot etalon with effective index of refraction $n_{eff}$.
The transmitted wavelength is given by
\begin{equation}
\lambda=\lambda_{max}\sqrt{1-\frac{\sin^2\theta}{n_{eff}^2}}
\label{tunability}
\end{equation}
where $\theta$ is again the angle of incidence. $\lambda_{max}$ is
the wavelength at normal incidence. A typical value is $n_{eff}=2$
(see section \ref{sectionFilter}). We choose the nominal
wavelength to be transmitted at $6^\circ$ of incidence which, for
$\lambda_{max}=853$~nm leads to
\begin{equation}
\frac{d\lambda}{d\theta}=-23\;\text{pm/mrad.}
\label{dlambdadthetaFilter}
\end{equation}
This is 60 times smaller than for the Littrow configuration. The
corresponding reduction of the sensitivity of the wavelength against
mechanical instabilities is a clear advantage of our design. Note
that this reduction is not achieved at the expense of a reduced
tunability (see section \ref{section698}). Tuning the filter
displaces the output beam by
\begin{equation}
\frac{dx}{d\lambda}=8\;\mu\text{m}/\text{nm,} \label{displFilter}
\end{equation}
due to the 0.5~mm thick fused-silica filter substrate ($n\sim
1.45$). This is two times smaller than for the modified Littrow
laser. Note that if one chooses to out-couple from the diode's back
facet, the displacement is further reduced.

\section{Sensitivity of the optical feedback}

We now study the sensitivity of the laser to a misalignment of the
external cavity that does not affect the emission wavelength but the
optical feedback. We consider a Gaussian beam with electric field
$E_{do}$ being emitted from the output facet of the diode ($z=0$).
Its propagation through the external cavity (along $z$) can be
modeled in the paraxial approximation giving the reflected electric
field $E_{dr}$ at the output facet of the diode. The feedback
$F=R^{-1}|\iint E_{do}^*E_{dr}dxdy|^2$ is given by the overlap
integral of the reflected and emitted electric fields. For
convenience we normalize by the reflectivity $R$ of the
grating/out-coupler. The variation of $F$ under misalignment
reflects the mechanical and thermal sensitivity of the laser.

Two sources of misalignment are considered: tilt of the out-coupler
(grating) and axial displacement of the out-coupler. Displacements
of other optical elements can be transformed into one of these. The
computation of $F$ turns out to be independent of the number of
lenses in the cavity, and can be simplified by calculating the
overlap integral $F=R^{-1}|\iint E_{rei}^*E_{rer}dxdy|^2$ at the
position $z=z_{re}$ of the reflective element \cite{OpticsLetters}.
Here $E_{rei}$ and $E_{rer}$ are respectively the incident and
reflected electric fields on the out-coupler. We assume that the
incident beam is perfectly aligned so that it forms a waist of
$1/e^2$ radius $w_0$ on the out-coupler.

If $\alpha$ is the angle formed by the incident and reflected beam
due to a small tilt of the out-coupler, we find \cite{OpticsLetters}
\begin{equation}
F=\exp\left[-(\alpha \pi w_0/\lambda)^2\right]\label{tilt}
\end{equation}
and for $\alpha\rightarrow0$
\begin{equation}
\frac{\partial^2F}{\partial\alpha^2}=-\frac{2\pi^2 w_0^2}{\lambda^2}
\label{derivtilt}
\end{equation}
Note that  $\xi=\lambda/(\pi w_0)$ is the $1/e$ divergence angle.

On the other hand, if the reflective element is displaced along
the optical axis by $\delta$, the reflected beam has a radius of
curvature $r=2\delta+z_R^2/(2\delta)$ and $1/e^2$ radius
$w=w_0\sqrt{1+(z_R/(2\delta))^2}$ with Rayleigh length $z_R=\pi
w_0^2/\lambda$. This gives
\begin{equation}
F=\left(1+\frac{\delta^2\lambda^2}{\pi^2w_0^4}\right)^{-1}
\end{equation}
and for $\delta\rightarrow0$
\begin{equation}
\frac{\partial^2F}{\partial\delta^2}=-\frac{2\lambda^2}{\pi^2w_0^4}
\label{displacement}
\end{equation}

Equations \ref{derivtilt} and \ref{displacement} show that $w_0$ is
the only parameter which determines the sensitivity of the optical
feedback to misalignment. In grating-tuned extended cavity lasers,
the beam waist is essentially determined by the selectivity
requirement and is of the order of 1\,mm. This leads to a rather
poor trade-off between angular and displacement sensitivity. Indeed,
a tilt of the grating of $\alpha=100\,\mu$rad is sufficient to
decrease the coupling factor $F$ by $10\,$\%, while a similar
reduction due to a pure displacement would correspond to
$\delta=1\;$m. In the new scheme described here the separation of
the wavelength selection and optical feedback allows to choose a
more favorable value for $w_0$. In our cat's eye setup
$w_0\sim\,10\,\mu$m. Hence, the tilt or displacement reducing $F$ by
10~\% ($\alpha=9\,$mrad or $\delta=0.1\,$mm respectively) are both
very large deformations.

\section{The filter}\label{sectionFilter}

The interference filter \cite{REO} is formed of a series of
dielectric coatings on an optical substrate with anti-reflection
coated back face. It is calculated to transmit more than 90~\% of
the intensity at the nominal wavelength at $6^\circ$ incidence. The
fullwidth at half maximum (FWHM) of the transmission curve is chosen
as 0.3~nm, which is about twice the mode spacing of a typical laser
diode. Filters with even higher finesse can be produced only at the
cost of reduced transmission. The chosen compromise turns out to
provide sufficient discrimination for stable single mode lasing with
satisfactory output power (see section \ref{laser852}).

The samples tested here are fabricated on larger optical wafers and
then cut into pieces of $5\times5~\text{mm}^2$, thereby reducing the
production costs. We test several fabrication batches at 698, 780
and 852~nm nominal wavelength. In the following the measurements on
a 852~nm filter are described. Similar results are obtained for the
other wavelengths.

\begin{figure}[t]
\centerline{\rotatebox{-90}{\includegraphics[width=8cm]{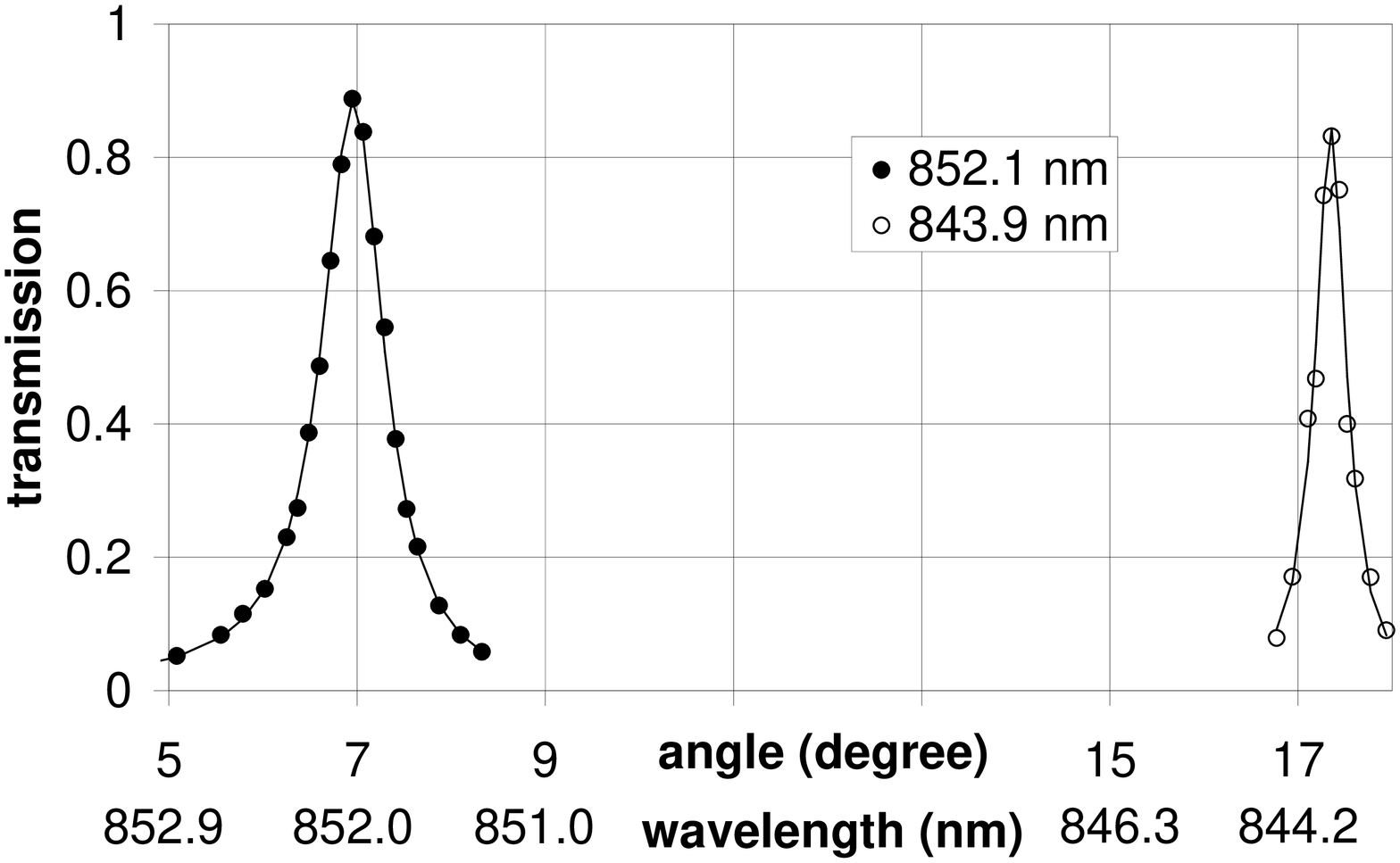}}}
\caption{Filter transmission of 852.1~nm light ($\bullet$) and
843.9~nm light ($\circ$), as a function of the angle of incidence.
The solid lines are Lorentzian fits to each peak.
\label{FinessFilter}}
\end{figure}

The transmission of a 1.2~mm diameter collimated beam of known
wavelength is measured. Fig. \ref{FinessFilter} shows the results as
a function of the angle of incidence for a 852.1~nm beam ($\bullet$)
and a 843.9~nm beam ($\circ$). The first maximum has 89~\%
transmission at $\theta=6.96(4)^{\circ}$ and a FWHM of
$0.80(1)^{\circ}$. The 843.9~nm light is transmitted to 84~\% at
$\theta=17.34^{\circ}$ with a FWHM of $0.40^{\circ}$. Using equation
\ref{tunability} we fit
 $\lambda_{max}=853.7$~nm and $n_{eff}=1.97$. This leads to
$\Delta\lambda_{FWHM}=0.37$~nm for 852~nm and
$\Delta\lambda_{FWHM}=0.44$~nm for 844~nm. The transmission peaks
are well fitted by a Lorentzian taking the wavelength as argument.
If the filter is used at a wavelength 8~nm below its nominal value
the transmission drops by 5~\% only. Repeating the measurement on 6
different production batches, we find less than 5~\% variation of
the maximum transmission at 852.1~nm.

%The high transmission, narrow resonance and easy tunability make
%this filter an ideal element for frequency selection in external
%cavity lasers.

In a second set of measurements, the transmission at 852.1~nm is
determined as a function of the incident polarization. A sinusoidal
variation from 89~\% to 75~\% is observed when the linear
polarization rotates from parallel to perpendicular with respect to
the axis of inclination.

\begin{figure}[t]
\centerline{\rotatebox{-90}{\includegraphics[width=8cm]{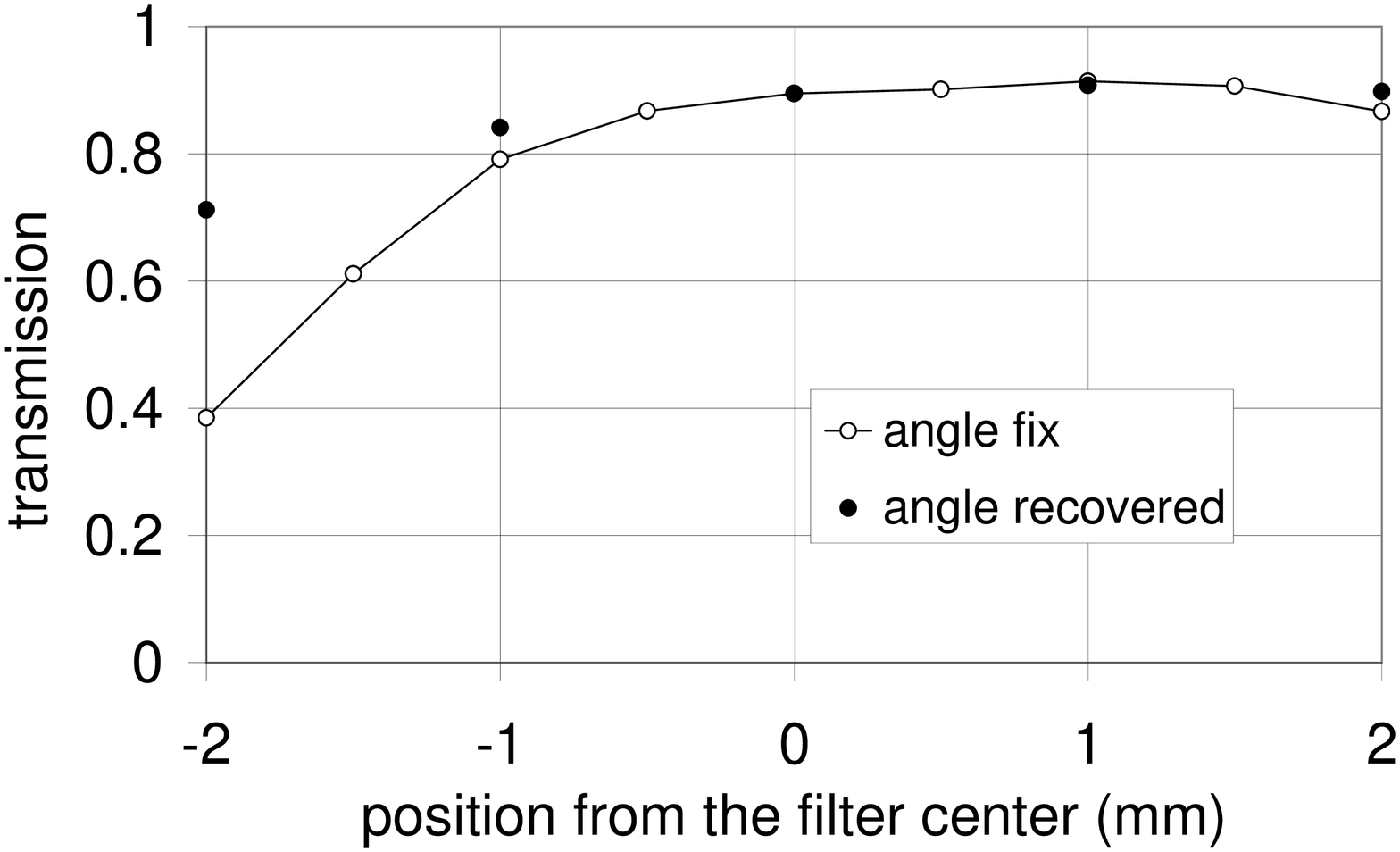}}}
\caption{Transmission of the 852~nm filter vs. position. When the
angle of incidence is kept constant, the transmission drops at the
left edge. If the angle is optimized at each point, it can be
recovered to $>70$~\%.\label{PosFilter}}
\end{figure}

Thirdly, the transmission is analyzed as a function of the position
on the filter (Fig. \ref{PosFilter}). A 0.5 mm wide slit is placed
in the 852.1~nm beam. Having optimized the angle of incidence at the
center and keeping it fixed ($\circ$), the right half of the filter
shows good homogeneous transmission, whereas the transmission drops
drastically towards the left edge. When the angle of incidence is
optimized at each position ($\bullet$), the transmission can be
recovered to $>~70$~\%. The corresponding angle at the left edge is
0.6 mrad bigger than the optimum angle at the center. Fig.
\ref{PosFilter} represents a batch with good spatial homogeneity.
Stronger variations are observed for other samples. This result
indicates that smaller beam diameters ($<1$~mm) are favorable.

Finally we test the filters under vacuum. At a residual pressure of
$10^{-4}$~Pa, we do not observe any variation of the optimal angle
of transmission compared to atmospheric pressure to within our
experimental resolution (5~GHz when expressed in terms of
frequency). To ensure that desorption of residual gas from the
coatings does not influence this result, the vacuum is kept over one
month. This makes the filter a perfectly suited candidate for
wavelength selection in space-born lasers.

%We have also tested the vacuum compatibility of the present filter,
%which has a particular importance for the design of space based
%extended cavity lasers. From atmosphere down to 10^-6 mbar we do not
%observe a significant variation of the maximum transmission. Keeping
%the filter at this low pressure during one month does not induce any
%changes either.

\section{A prototype emitting at 852 nm}\label{laser852}

Following the design of Fig. \ref{schematic}, a laser at 852~nm is
built. The diode (SDL 5422) nominally emits 150~mW for a current of
$I=150$~mA at 854~nm. The output facet coating is specified to
induce less than 4~\% reflection. We measure the free spectral range
of the naked diode as $\sim 50$~GHz (0.13 nm), which corresponds to
a physical length of $\sim0.8$~mm assuming the index of refraction
of GaAs ($n=3.6$) \cite{Siegman}.

The collimating lens has a focal length of 4.5~mm. L1 and L2 have
focal length $f=18.5$ and 11~mm, respectively. The out-coupler is an
optical flat of 10~mm diameter and 3~mm thickness. It is coated
partially reflective on the inner face and anti-reflection on the
outer. It is glued onto a 10~mm diameter PZT tube of 1~mm wall
thickness and 10~mm length. The overall length of the external
cavity is 70~mm.

Different reflection coefficients of the OC are tried. The output
power at $I=85$~mA is  47, 40 and 30 mW for 15, 20 and 30~\%
reflectivity, respectively. At 15 and 20~\% reflection, single mode
operation is obtained for certain intervals of the diode current
only. At 30~\% reflection stable single mode lasing is assured from
the threshold current ($\sim10$~mA) to 8 times its value. Operation
on the same diode mode is assured within a span of 4~mrad of filter
inclination (44~GHz), before an adjacent mode of the diode is
selected. This demonstrates the small sensitivity of our design to a
tilt of the selective element (equation \ref{dlambdadthetaFilter}).

\begin{figure}[t]
\centerline{\rotatebox{-90}{\includegraphics[width=10cm]{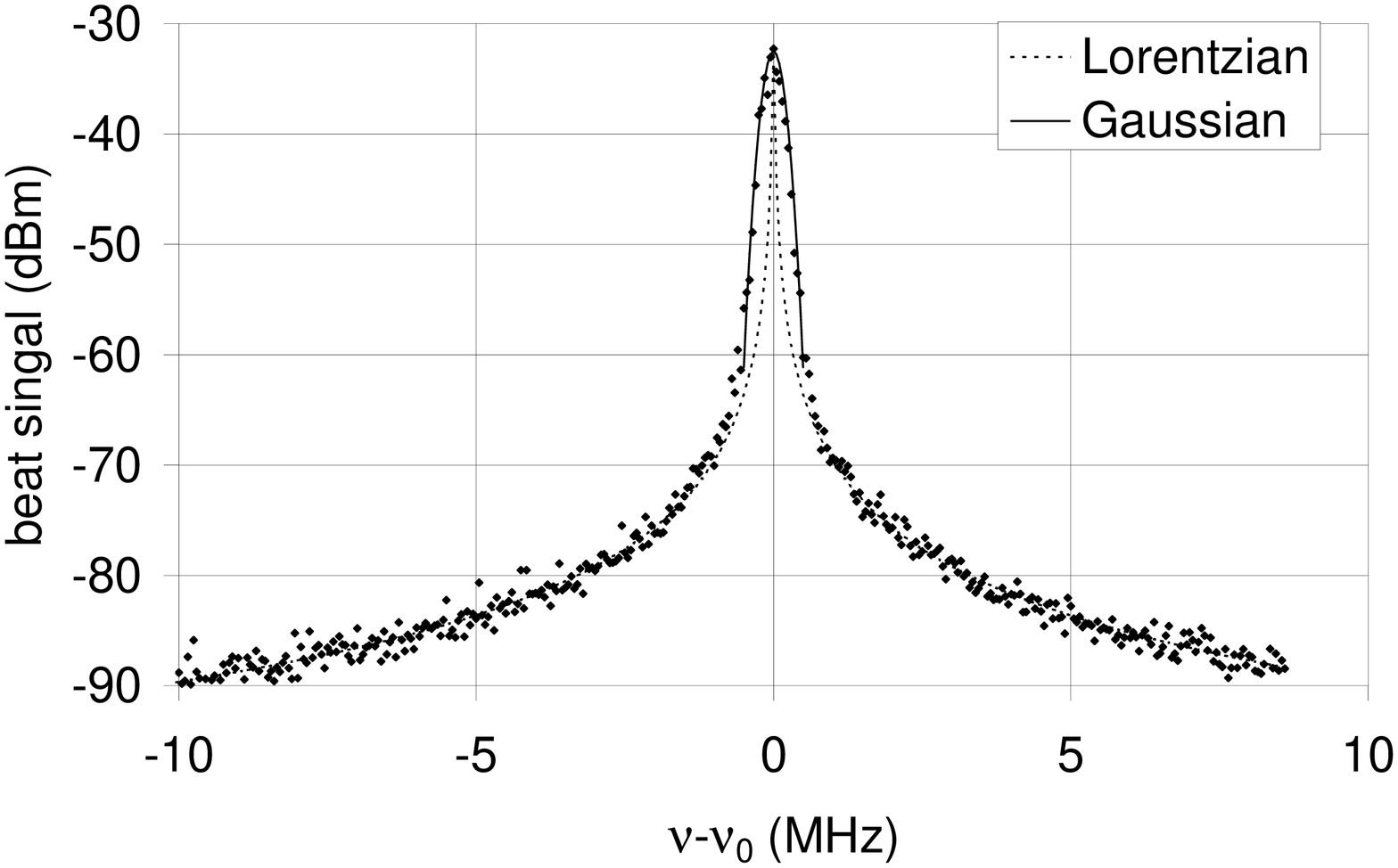}}}
\caption{Power spectrum of the beat signal at $\nu_0=8.75$~GHz
between two identical 852~nm lasers. The solid line is a Gaussian
fit to the central peak, the dashed line a Lorentzian fit to the
wings. The spectrum analyzer's resolution bandwidth is set to 1~kHz.
The full trace is swept in 25~s. \label{batement}}
\end{figure}

In order to determine the laser's spectral properties, we measure
the beat signal between two identical set-ups separated in frequency
by $\nu_0=8.75$~GHz. The observed spectrum is shown in Fig.
\ref{batement}. The first laser is locked to the D2 line of Cs. The
second one is locked (with a bandwidth of 15~kHz) to the first one
so that the beat frequency stays constant. The remaining higher
frequency noise leads to a random distribution of the central beat
frequency, which is well fitted by a Gaussian (solid line). We fit
the points within $\pm0.5$~MHz from the center and find a FWHM of
155~kHz. The wings ($|\nu-\nu_0|>1$~MHz) of the beat signal can be
fitted by a Lorentzian (dashed line) giving the high frequency noise
of the laser. Its FWHM of 28~kHz indicates that the white noise
level of each laser corresponds to a linewidth of 14~kHz. An even
smaller white noise floor may be obtained by means of stronger
optical feedback or a longer external cavity.

Because of the filter's vacuum compatibility and the mechanical
stability of our prototype, the design of Fig. \ref{schematic} has
been selected for the construction of space-qualified lasers for the
PHARAO project\,\cite{Salomon01}. Space qualification includes
survival in vibratory environments with sinusoidal excitation
(30~Hz) at a level of $\pm35~g$ during 120~s and random excitation
(20-600 Hz) with a level of 31 $g$ rms during 120 s. The
qualification model has passed the vibration tests. Its spectral
characteristics did not change and the output beam mis-orientation
remained below 10 $\mu$rad.

Using a \emph{Sharp} diode (GH0781JA2C) and a 780~nm filter in a
similar external cavity we generate laser light for the manipulation
of atomic rubidium.

\section{A prototype emitting at 698 nm}\label{section698}

Similar components are used to build a laser at 698 nm with a 100~mm
long external cavity. A \emph{CircuLaser} diode (PS 107) specified
at 688 nm is used. Its free running output power is 30~mW for
$I=100$~mA at room temperature. The interference filter used here is
specified for 698 nm at 6$^{\circ}$ of incidence. The measured
optimal incidence is 9$^{\circ}$. By varying the inclination of the
filter, the temperature of the diode and the reflection coefficient
of the OC, we achieve lasing from 679 nm to 700 nm with the same non
AR-coated diode. Emission at 679 nm is easily obtained by using a
40~\% reflection OC and cooling the diode to 19$^{\circ}$C. The
output power is 4 mW, for $I=60$~mA. Even smaller wavelengths seem
reachable, only limited by the size of the filter mount. It is more
difficult to reach 698 nm, as this wavelength lies at the very edge
of the diode's gain curve. We heat the diode to 40$^{\circ}$ C.
60~\% reflectivity of the OC gives stable single mode lasing but
limited output power. 50~\% reflection leads to an output power of
2~mW at $I=57$~mA with sufficient mode selection. No emission at
698~nm is observed for 40~\% reflection. Finally, due to operation
at the very edge of the gain curve, the possibility to tune the
laser by varying the current of the diode laser is limited to
40\,GHz, as compared to 70 GHz if the diode is emitting at its
nominal wavelength.

The tunability of this laser is significantly larger than we had
 formerly achieved with grating feedback. This may be due to the fact
  that the coupling of the diode to the external cavity is more easily optimized with
the design of figure \ref{schematic}. In addition, the grating
induces aberrations which are difficult to compensate for. Being
able to pull the diodes more than 10~nm away from their nominal
wavelength is another advantage of the new setup, especially in
spectral regions where laser diodes are not available, like 700\,nm.

%The tunability of this laser is significantly larger than we had
% formerly achieved with grating feedback. This is probably due to the fact that the
%optimization of the coupling of the diode to the external cavity is
%much easier with the design of figure \ref{schematic}. In addition,
%the grating induces aberrations which are difficult to compensate
%for. In previous experiments in Littrow configuration, 698 nm was
%the very upper limit that we could reach. This was done by adjusting
%the laser polarisation so as to maximize the optical feedback by a
%blazed grating. The diode also had to be heated up to 60$^{\circ}$C.
%The output power was limited to a few hundreds of $\mu$W. Being able
%to pull the diodes further away from their nominal wavelength is
%certainly another advantage of the new setup, especially at the edge
%of spectral regions where laser diodes are not available, like
%700\,nm.

\begin{figure}[t]
\centerline{\includegraphics{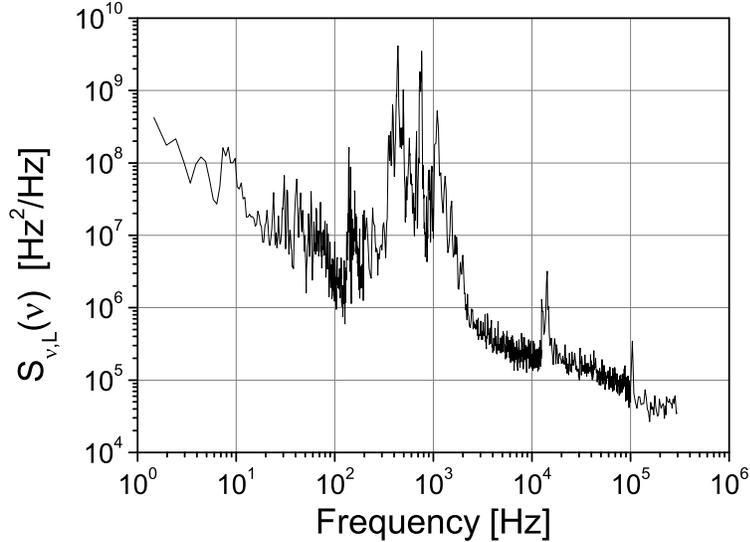}} \caption{Power spectal
density of frequency fluctuations of the 698~nm prototype.
\label{noise}}
\end{figure}

We measure the frequency noise of the interference filter laser
against a 400 mm long Fabry-P\'erot cavity of finesse 300. Both, the
laser and the cavity are free running. The observed frequency noise
power spectral density $S_{\nu,L}(\nu)$ is plotted in Fig.
\ref{noise}. From 2~Hz to 300~kHz, $S_{\nu,L}(\nu)$ essentially
decreases with $1/\nu$ corresponding to flicker frequency noise.
Peaks due to acoustic perturbations are also observed for Fourier
frequencies between 100~Hz and 2~kHz. From Fig. \ref{noise} we can
deduce the fast linewidth $\Delta\nu_{L}$ of the laser using
\cite{linewidth}
\begin{equation}\int_{\Delta\nu_{L}/2}^{\infty}S_{\phi,L}(\nu)d\nu
=\frac{2}{\pi}
\end{equation}
where $S_{\phi,L}=S_{\nu,L}/\nu^{2}$ is the power spectral density
of phase fluctuations. We find a fast linewidth
$\Delta\nu_{L}=150$~kHz.

\section{Conclusion}

We have built external cavity diode lasers using an interference
filter as wavelength selective element. The filter presents high
transmission and narrow bandwidth. The cavity design drastically
improves the laser's passive stability and reduces beam walk when
tuned. The amount of feedback can easily be varied by the
reflectivity of the out-coupler. Prototypes at different wavelength
in the visible and near infra-red were realized showing a linewidth
down to 14~kHz. Tuning over 20~nm was demonstrated. The prototypes
are currently employed in atomic physics experiments on Cs, Rb and
Sr, e.g. for laser cooling.

%After the manuscript is proofread, the {\tt .tex} file and figures
%should be archived with tar-gzip compression. Do not include
%subdirectories within the archive.

%To upload your manuscript, follow the instructions on the each
%journal's homepage (see \mbox{\href{http://www.opticsinfobase.org}{http://www.opticsinfobase.org}}).
%Authors should feel free to contact OSA staff for assistance; details are available at \href{InfoBase}{http://www.opticsinfobase.org}.

%\appendix

%\section*{Appendix A: Sample}
%\setcounter{equation}{0}
%\renewcommand{\theequation}{A{\arabic{equation}}}

%\begin{equation}
%a+b=c.
%\end{equation}

\section*{Acknowledgements}
We are grateful to A. Landragin for fruitful discussion.


\begin{thebibliography}{99}
\bibitem{indexOfRefraction} Since a grating selects the wavelength in air,
passing from ambient pressure to vacuum changes the index of
refraction by $3\times10^{-4}$ corresponding to
$\Delta\lambda=0.26$~nm (100 GHz) at $\lambda=852$~nm.
\end{thebibliography}
\end{document}